# Discovering novel drug-supplement interactions using a dietary supplements knowledge graph generated from the biomedical literature

**Dalton Schutte[1,2], Jake Vasilakes[1,2,5], Anu Bompelli[2], Yuqi Zhou[1,2], Marcelo Fiszman[6], Hua Xu[7], Halil Kilicoglu[4], Jeffrey R. Bishop[3], Terrence Adam[1,2], Rui Zhang[1,2,*]**

[1] Institute for Health Informatics, University of Minnesota, Minneapolis, Minnesota, USA
[2] Department of Pharmaceutical Care & Health Systems, University of Minnesota, Minneapolis, Minnesota, USA
[3] Department of Experimental and Clinical Pharmacy, University of Minnesota, Minneapolis, Minnesota, USA
[4] School of Information Sciences, University of Illinois, Champaign, Illinois, USA
[5] National Centre for Text Mining, School of Computer Science, The University of Manchester, Manchester, United Kingdom
[6] NITES - Núcleo de Inovação e Tecnologia Em Saúde, Pontifical Catholic University of Rio de Janeiro, Brazil
[7] School of Biomedical Informatics, The University of Texas Health Science Center at Houston, Houston, Texas, USA

*Corresponding Author Contact:
Rui Zhang, PhD
8-100 Phillips-Wangensteen Building
516 Delaware Street SE
Minneapolis, MN 55455
University of Minnesota
zhan1386@umn.edu

**Abstract**

OBJECTIVE: Leverage existing biomedical NLP tools and DS domain terminology to produce a novel and comprehensive knowledge graph containing dietary supplement (DS) information for discovering interactions between DS and drugs, or Drug-Supplement Interactions (DSI).

MATERIALS AND METHODS: We created SemRepDS (an extension of SemRep), capable of extracting semantic relations from abstracts by leveraging a DS-specific terminology (iDISK) containing 28,884 DS terms not found in the UMLS. PubMed abstracts were processed using SemRepDS to generate semantic relations, which were then filtered using a PubMedBERT-based model to remove incorrect relations before generating our knowledge graph (SuppKG). Two pathways are used to identify potential DS-Drug interactions which are then evaluated by medical professionals for mechanistic plausibility.

RESULTS: Comparison analysis found that SemRepDS returned 206.9% more DS relations and 158.5% more DS entities than SemRep. The fine-tuned BERT model obtained an F1 score of 0.8605 and removed 43.86% of the relations, improving the precision of the relations by 26.4% compared to pre-filtering. SuppKG consists of 2,928 DS-specific nodes. Manual review of findings identified 44 (88%) proposed DS-Gene-Drug and 32 (64%) proposed DS-Gene1-Function-Gene2-Drug pathways to be mechanistically plausible.

DISCUSSION: The additional relations extracted using SemRepDS generated SuppKG that was used to find plausible DSI not found in the current literature. By the nature of the SuppKG, these interactions are unlikely to have been found using SemRep without the expanded DS terminology.

CONCLUSION: We successfully extend SemRep to include DS information and produce SuppKG which can be used to find potential DS-Drug interactions.

**1. INTRODUCTION**

The 1994 Dietary Supplement Health and Education Act (DSHEA) defines a dietary supplement (DS) as a product intended to supplement the diet that contains one or more of the following dietary ingredients: a vitamin, a mineral, an herb or other botanical, an amino acid, a dietary substance to supplement the diet or a concentrate, metabolite, constituent, extract, or combination of any ingredients thereof[1]. Data from the 2017-2018 National Health and Nutrition Examination Survey (NHANES) found that 57.6% of U.S. adults aged over 20 used some sort of DS. The data also shows that across the age groups for both men and women, prevalence of DS use increases with age[2]. The DSHEA also classifies DSs as a category of food and thus do not require pre-market approval by the FDA like pharmaceutical drugs. Furthermore, many herbs are non-patentable so there is less incentive to conduct research to clarify their safety, efficacy, and drug interaction potential[3].

This leads to the effects of many DS not being adequately understood. With this comes the risk that an individual may experience an interaction between a DS they are taking and a pharmaceutical substance. Some systematic reviews have investigated the literature for interactions between drugs and DSs[4–9]. However, they note the limited literature studying Drug-Supplement Interactions (DSI), the quality of the reviewed studies, the sample size in the reviewed studies, etc. as being limitations. The literature suggests that there is still insufficient knowledge surrounding DS and their potential interactions with other substances, adverse or otherwise.

To address these limitations, we created an extended version of SemRep, called SemRepDS, by integrating more DS terms from a DS specific knowledge base with the UMLS to enable better recognition of DS terms in biomedical literature. We then produced a unique knowledge graph, SuppKG, containing many relationships found in SemMedDB as well as many unique relationships. Finally, we utilized SuppKG to identify novel DSIs through literature-based discovery.

## 2. BACKGROUND AND RELATED WORK

### 2.1 SemRep and SemMedDB

MetaMap[10,11] links entity mentions in biomedical text to concepts in the UMLS Metathesaurus. It returns a mapping decision for each identified phrase in a body of text. SemRep is a rule-based tool for extracting semantic relations from biomedical text using MetaMap to map text to concepts in the UMLS [12]. Applying SemRep to the entire collection of PubMed citations produces a large database of relations, entities, and source sentences called SemMedDB[13]. While an extension of SemMedDB has been produced before (SemMedDB UTH[14]) this method involved extending SemRep by modifying the Metathesaurus directly to undo the suppression of the drugs in the National Cancer Institute Thesaurus. Prior work has demonstrated that SemRep can be extended to the biomedical domain of disaster medicine [15]and medical informatics [16]using various computational techniques.

### 2.2 Literature Based Discovery

Literature Based Discovery (LBD) is a method for generating hypotheses by finding implicit relationships in the research literature[17]. Two categories of LBD are open and closed LBD. Open LBD involves providing a term and the task is to find connections between two concepts that may not be directly related. Closed LBD takes two terms and finds concepts shared between them[18]. Discovery patterns exploit the explicit relationships between UMLS concepts by imposing conditions on what pathways are valid, potential relationships[17].

Previous work on Drug-Drug Interactions (DDIs) discovery has used a variety of data sources and techniques. One study identified combinations of Drug-Gene relationships associated with known DDIs using Medline abstracts and a random forest classifier[19]. Another used electronic medical records (EMR) to narrow a set of potential DDIs generated from *in vitro* studies to a collection of drug pairs that were found to have higher risk ratios for myopathy[20]. SemMedDB [21] has also been used in the past to discover potential DDIs using discovery patterns based on semantic types of the UMLS concepts extracted from PubMed abstracts in combination with patient EMR data[22].

In a recent study that performed LBD for drug repurposing[23], a BERT [24] model was fine-tuned on annotated semantic relations to identify which relations were correctly implied by their source sentence. The fine-tuned BERT model was used to remove incorrect relations from SemMedDB before open LBD was performed for drug repurposing.

### 2.3 The Integrated Dietary Supplement Knowledge base (iDISK)

The Integrated Dietary Supplement Knowledge base (iDISK) is a data model that contains terminology of DS ingredients from a collection of sources[25]. Like the UMLS data model, the core iDISK data elements are atoms, concepts, concept attributes, relationships, and relationship attributes. Included in the iDISK data model are links to other controlled vocabularies to allow for wider use and integration with existing biomedical knowledge representations such as the UMLS. iDISK contains 144,654 unique concepts, of which 4,208 are concepts for DS ingredients and 137,568 concepts for DS products.

### 2.4 Drug Supplement Interactions

There are limited studies that use LBD to discover DSIs. Our group previously discovered interactions between cancer drugs and DS through LBD[26]. We leveraged SemMedDB to identify both known and unknown DSIs through expert validation. Recently, the Allen AI Institute conducted DSI identification using articles from Semantic Scholar and RoBERTa[27,28]. Supp.AI mined the literature to find published DSIs. The Supp.AI lexicon of DS is based on data mined from Medline articles that have UMLS CUIs. Our prior work has shown that the UMLS coverage for DS terminology and synonyms is incomplete [29] which imposes an inherent limitation on the terminology Supp.AI can use. Comparing terms available in iDISK [25] and in Supp.AI we found that 87.3% of the concepts in Supp.AI were contained in iDISK but only 43.5% of the concepts in iDISK were contained in Supp.AI.

## 3. MATERIALS AND METHODS

We use the terminology and relations contained in iDISK to extend the domain of MetaMap and have SemRep use this extended version of MetaMap. A collection of abstracts from PubMed was processed by SemRepDS to extract semantic relations. A biomedical BERT variant was fine-tuned to filter out incorrect semantic relations from the dataset. Discovery patterns were used on the resulting knowledge graph to find DSIs. Figure 1 provides a visual summary of the entire process.

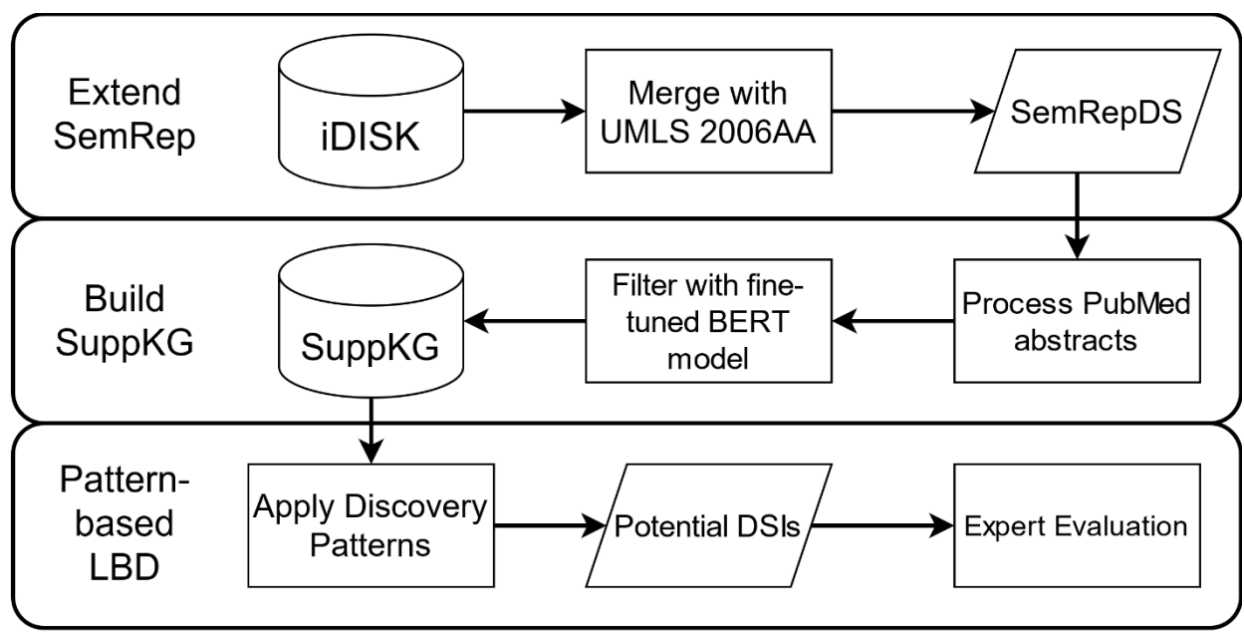

**Figure 1.** Overview of the methodology.

### 3.1 Integrating iDISK into UMLS Metathesaurus to create SemRepDS

The iDISK ingredient terminology was added to UMLS Metathesaurus to increase the DS representation. To achieve this, the MetaMap Datafile Builder[1] pipeline was used to generate the necessary data files required for MetaMap [10,11] which were then mapped to the DS extended Metathesaurus. The extended Metathesaurus was then linked to SemRep for relation extraction, henceforth, SemRepDS.

We extend the UMLS 2006AA Metathesaurus, since SemRep was optimized for this version, with DS ingredient concepts from iDISK as follows. For each iDISK concept with no existing links to any UMLS concept, we created a new concept entry in the UMLS MRCONSO.RRF file. This involved 1) defining a new concept unique identifier (CUI) for the DS concept, and 2) writing all the concept's atoms to MRCONSO.RRF. The UMLS defines preferred terms for each concept, which specify the canonical name for the concept, and which are determined according to a source vocabulary ranking defined in MRRANK.RRF. iDISK also specifies preferred terms for each concept according to a source ranking, and we designated atoms as "preferred" accordingly for all new DS concepts added to the UMLS. Adding iDISK concepts that *do* have existing links to the UMLS is a similar process, but we used the CUI of the linked UMLS concept instead of creating a new one, and we kept the existing UMLS preferred term rather than using the iDISK preferred term. For all iDISK concepts added, we ensured that they were assigned the "Pharmacologic Substance (phsu)" semantic type, as this semantic type is used by SemRep to determine which concepts are potential candidates for our target semantic relations, INTERACTS_WITH, INHIBITS, AUGMENTS, etc. The semantic types for each concept were written to MRSTY.RRF. We also added entries for the iDISK source vocabularies to MRSAB.RRF and MRRANK.RRF. Finally, the RRF files were fed into the MetaMap Data File Builder, which indexed the files and generated a concept database that can be used by MetaMap.

---

[1] https://metamap.nlm.nih.gov/DataFileBuilder.shtml

We selected English PubMed abstracts using the terms contained in iDISK as search terms, resulting in 608,725 unique PMIDs. To compare the ability of SemRepDS for extracting DS entities and semantic relations, we processed the abstracts separately using both base SemRep v1.8 and SemRepDS (mentioned in 2.1).

### 3.2 Filtering and ranking semantic relations with PubMedBERT

To improve the quality of the knowledge graph, relations were filtered for correctness using a BERT [24] model that had been fine-tuned for the binary classification task using 6000 annotated relations (inter-annotator agreement=0.842)[30], developed previously[23]. To further include semantic relations with DS mentions, we randomly collected 300 abstracts containing 492 relations with DS mentions for annotation by two informatics graduate students with backgrounds in pharmacy and pharmaceutical sciences (AB and YZ). Each relation was labeled as correct or incorrect, with a "correct" relation indicating that the extracted relation was verified as included in the source sentence. Among the combined annotated dataset, 67.02% (4,251/6,492) were labeled as correct relations.

The combined dataset was split 70/20/10 for training, development, and test sets, respectively. The train and development sets were sampled so that they were balanced evenly between correct and incorrect labelled relations. The test set was sampled to match the original distribution of correct/incorrect relations.. Six BERT variants, including base uncased, PubMedBERT abstracts only, PubMedBERT abstracts and full text [31], BioBERT [32], BlueBERT [33], and BioClinicalBERT [34], were trained and evaluated on the dataset seven times. Hyperparameters for the training were: five epochs, learning rate of 0.0003, weight decay of 0.1, gradient clipping was used with a max of 1, batch size was 16, and 200 warmup steps were used at the beginning of each training cycle. Optimization was performed using Adam [35] with decoupled weight decay and a cosine learning rate decay. The PubMedBERT variant obtained the best mean F1-score and was used for filtering.

### 3.3 Evaluating semantic relations generated by SemRepDS

To evaluate the accuracy of the process, we used SemRepDS to process 300 randomly selected abstracts followed by human evaluation to identify any potential systematics errors. Since the relations from the 300 abstracts were used to train the BERT filter, we also sampled 50 abstracts containing 224 relations before filtering. The same informatics students (AB and YZ) assessed the correctness of these relations based on the source sentences. We then report the precision of semantic relations before and after filtering.

### 3.4 Creating SemMedDB and SuppKG

This filtered collection of relations constitutes SemMedDB-DS. To incorporate more generic, non-DS specific terms for our LBD, we augmented SemMedDB-DS with a collection of relations from a previous study [23]. In that study, SemMedDB 43 augmented with additional relations generated by using SemRep on the CORD-19 dataset was filtered by selecting relations with particular semantic types and predicates useful for drug-repurposing, and after additional processing, reduced the number of relations from over 108 million to 2.5 million. We added the remaining 2.5 million relations from the prior study [23]to our 2 million relations, since the semantic types and predicates useful for drug repurposing are also useful for DSI discovery. This resulted in a total of 4,521,483 relations. Our dataset combined with the additional dataset contain relations from 2,035,460 unique abstracts. These were then filtered using our trained PubMedBERT model.

This collection of relations can be represented using a graph-structured data model where the subjects and objects form the set of graph nodes, *N*, and the predicates form the set of directed edges, *E*. We thus created a knowledge graph, called SuppKG, using only the filtered set of relations from SemRepDS. For DSI discovery, we combined SuppKG and the relations from the prior study [23]. Our combined contains graph contains 130,763 nodes with 1,434,007 directed edges. SuppKG alone contains 56,635 nodes with

595,222 directed edges. SuppKG will be made available as a NetworkX graph object stored in a pickled file and a json file[2].

### 3.5 Discovery and evaluation of DS-Drug Interactions (DSI)

Two discovery pathways were used for interaction discovery: DS-Gene-Drug (DsGD) and DS-GeneA-Biological Function-GeneB-Drug (DsGFGD), similar to our prior study [22]. For each pathway, the interactions were ranked in descending order based on the sum of each semantic relations' accuracy confidence scores assigned by the classifier.

The evaluation of the top 50 DSI pathways from each pattern was conducted by two of the authors (TA and JB), who are pharmacists with pharmaceutical sciences and clinical backgrounds. A protocol was developed to operationalize a review and evaluation process and ensure similar resources were used by both experts. The first step in evaluating a pathway was to identify all terms. This was accomplished by confirming that the text-to-concept mapping was correct then checking the UMLS, PubMed, and, as a final effort, an internet search to identify the terms. The text to concept mapping required verified since MetaMap does not always perfectly map text to the proper concept. The second step was to evaluate the relationships themselves by confirming the relations were correctly extracted from the abstract, checking the associated paper(s), and, finally, an internet search. If after these steps, they felt that the DSI suggested by the pathway was correctly extracted from the associated abstracts and if each relation in the pathway represented a logical biochemical/microbiological connection, they rated the DSI as 'plausible'. Otherwise, the DSI was determined to be 'implausible'. We then report the percentage of plausible DSIs and discuss some examples.

To check if our DSI approach identified known DSIs, we compared our list against the Natural Medicine Interaction Checker database [36]. Natural medicine is expert-curated monographs for natural products, which include drug-supplement interactions.

## 4 RESULTS

### 4.1 Comparison between SemRep and SemRepDS

Each semantic relation was paired with its source sentence in a database for comparison and further processing. In this study, Table 1 demonstrates that SemRepDS increased the number of DS entities identified by 158.52% after adding the iDISK terminology with UMLS. The expanded DS entities further improved the identification of additional 148,308 (206.93%) relations with at least one DS entity

Table 1: Comparison of the output from SemRep and SemRepDS

|  | SemRep | SemRepDS | Difference |
| --- | --- | --- | --- |
| DS Entities Mentions | 539,863 | 1,395,653 | 855,790 (+158.52%) |
| Relations with at least one DS Entity | 71,669 | 219,977 | 148,308 (+206.93%) |

### 4.2 Performance comparison of semantic relation correctness classifiers.

Table 2 contains metrics from experiments testing various BERT models. BioBERT had the highest mean precision, PubMedBERT pre-trained on PubMed abstracts and full text articles the highest mean recall, and PubMedBERT pre-trained only on PubMed abstracts the highest mean $F_1$ score. Each model was trained and evaluated seven times to compute mean scores and confidence intervals.

---

[2] https://github.com/zhang-informatics/SemRep_DS/tree/main/SuppKG

Table 2: Performance for the evaluated BERT-based classifiers. Metrics represent the mean score across 7 training runs with upper and lower 95% confidence intervals (i.e., mean (lower CI, upper CI))

|  | Precision | Recall | F1 |
| --- | --- | --- | --- |
| BERT | 0.8528 (0.8292, 0.8763) | 0.7451 (0.6810, 0.8093) | 0.7930 (0.7653, 0.8207) |
| BioClinicalBERT | 0.8147 (0.8020, 0.8273) | 0.7624 (0.7363, 0.7885) | 0.7872 (0.7767, 0.7977) |
| BLUEBERT | 0.8666 (0.8484, 0.8848) | 0.7654 (0.7186, 0.8123) | 0.8115 (0.7911, 0.8320) |
| BioBERT | **0.8898** (0.8756, 0.9039) | 0.7636 (0.7052, 0.8220) | 0.8203 (0.7916, 0.8490) |
| PubMedBERT Abstracts + Full Text | 0.8489 (0.8257, 0.8721) | **0.8469** (0.7917, 0.9020) | 0.8461 (0.8284, 0.8638) |
| PubMedBERT Abstracts | 0.8796 (0.8539, 0.9053) | 0.8269 (0.7730, 0.8808) | **0.8506** (0.8317, 0.8695) |

### 4.3 Evaluation of semantic relations generated by SemRepDS

Among the 300 abstracts included based on random sampling, the precision of the pre-filtering SemRepDS output was found to be 0.67. This is comparable to a recent evaluation of SemRep that found a precision of 0.69 [37]. After filtering, PubMed BERT was then used to filter generated semantic relations down to 2,710,240 (59.94%) For the sample of 50 random abstracts, before filtering the precision was 0.72 and after filtering increased to 0.91.

### 4.4 Statistics of DS knowledge graph - SuppKG

Table 3: Distribution of the predicates after filtering the combined relations with PubMed BERT

| Predicate | Count | Predicate | Count |
| --- | --- | --- | --- |
| TREATS | 525,719 (19.40%) | ISA | 24,383 (0.90%) |
| COEXISTS_WITH | 511,108 (18.86%) | PREDISPOSES | 21,236 (0.78%) |
| PROCESS_OF | 257,484 (9.50%) | COMPARED_WITH | 19,295 (0.71%) |
| CAUSES | 235,599 (8.69%) | ADMINISTERED_TO | 18,940 (0.70%) |
| INTERACTS_WITH | 213,407 (7.87%) | METHOD_OF | 15,203 (0.56%) |
| AFFECTS | 209,644 (7.74%) | DIAGNOSES | 6,531 (0.24%) |
| LOCATION_OF | 169,137 (6.24%) | MEASURES | 4,562 (0.17%) |
| PART_OF | 131,192 (4.84%) | PRECEDES | 3,132 (0.12%) |
| ASSOCIATED_WITH | 120,297 (4.44%) | COMPLICATES | 1,846 (0.07%) |
| USES | 96,211 (3.55%) | HIGHER_THAN | 1,817 (0.07%) |
| INHIBITS | 59,524 (2.20%) | OCCURS_IN | 1,535 (0.06%) |
| AUGMENTS | 55,167 (2.04%) | MANIFESTATION_OF | 1,199 (0.04%) |
| DISRUPTS | 45,017 (1.66%) | CONVERTS_TO | 1,055 (0.04%) |
| PRODUCES | 41,402 (1.53%) | SAME_AS | 156 (0.01%) |
| STIMULATES | 39,332 (1.45%) | LOWER_THAN | 110 (0.00%) |
| PREVENTS | 39,104 (1.44%) |  |  |
| TOTAL |  | 2,710,240 |  |

### 4.5 Mechanistic Evaluation

The expert review found that 88% (44/50) of the DsGD relations and 64% (32/50) of the DsGFGD relations were mechanistically correct. The mechanistic correctness of a relation does not necessarily imply clinical utility and are considered 'clinically plausible'.

### 4.6 Comparison of existing DSI knowledge base

Among 100 DSI list, five DSIs found with the DGD pathway were found in the Natural Medicines [36] database and none of the DSIs found with the DGFGD pathway were found in the database. The known interactions are:

```
flaxseed INHIBITS tnf receptor ligands STIMULATES plasminogen activator
urokinase
curcumin INHIBITS interleukin 1, beta STIMULATES plasminogen activator
urokinase
allicin INHIBITS inerleukin 1, beta STIMULATES plasminogen activator
urokinase
activin INHIBITS inerleukin 1, beta STIMULATES plasminogen activator
urokinase
vanadium STIMULATES atp STIMULATES vasopressin
```

## 5 DISCUSSION

DSIs are recognized and have potential risks for patients, however interactions between DS and pharmaceutical drugs are neither widely understood nor well identified in the biomedical literature. One of such barriers is the incomplete representation of DS terminology in the current biomedical terminologies. [29]. Until now, there was no prior study that integrated a new, specialized vocabulary (i.e., iDISK in this study) with the UMLS and used the resulting extended UMLS for relation extraction. We successfully extended the UMLS with a DS-specific vocabulary and used the resulting extended UMLS for discovering substance interactions from the literature. We also successfully used a fine-tuned BERT model to filter incorrect semantic relations. As such, we demonstrated the utility of vocabulary extension by applying discovery pathways to SuppKG to find novel DSIs, most of which were determined to be mechanistically plausible.

### 5.1 SemRepDS and SemMedDB-DS

In our prior work, we found that iDISK can enrich UMLS to represent DS and can further improve performance on a NER task [29]. In this study, we further integrated iDISK terms with UMLS through MetaMap Data File Builder, which was demonstrated to increase the recall of recognizing DS terms and DS-related relations. This results in more relations generated by SemRepDS than SemRep. Compared with SemRep, SemRepDS can extract entities and relations with specific DS mentions, demonstrating SemRepDS' ability to extract data that would otherwise be absent. Thus, the additional relations extracted with SemRepDS are unique and not found in SemMedDB.

There are 49,571 additional relations that were extracted with SemRepDS which gives us a richer knowledge graph to work with. Furthermore, 512,201 (25.55%) of the relations extracted by SemRepDS contain at least one DS mention. Using a knowledge graph constructed from SemRepDS output contains more relations as well as relations with specialized terms that will facilitate DSI discovery.

Perhaps more importantly, we have demonstrated that specialized terminology can successfully be integrated with the UMLS Metathesaurus using the Data File Builder and the resulting extended Metathesaurus used to identify additional entities and relations. Extending the UMLS with a specialized terminology can be done with other domains beyond DS [15,16]. Additionally, the use-cases for an extended UMLS are not limited to DSI discovery. The extended UMLS was used for other types of literature-based discovery, information extraction, and other informatics tasks.

### 5.2 Filtering

A sample of 50 abstracts had a pre-filtering precision of 0.72 and a post-filtering precision of 0.91. This is a substantial improvement over the raw output from SemRepDS. While the resulting knowledge graph is smaller, the likelihood of any path being comprised of true relations is higher. The potential DSIs identified by our pathways are likely to be of a higher quality than if we used the same pathways on the pre-filtered graph.

It is somewhat surprising that the PubMedBERT model trained on abstracts only outperformed the one trained on abstracts and full-text articles. However, it was observed by the authors in [31] that the abstract-only out-performed the abstract+full-text model on some tasks. Since our dataset is derived from PubMed abstracts, the inclusion of full-text articles likely resulted in a shift away from the target distribution that hindered the full-text model compared to the abstract only model, which contains many of the abstracts included in our sample. An abstract contains more distilled statements regarding the exact nature of a study and its findings. Thus, the abstract-only model would have a more unadulterated representation of the relationships that SemRepDS is extracting. The PubMedBERT model used for filtering obtained an F1 score of 0.87 which suggests that the model was reasonably successful in identifying correct relations.

### 5.3 Link Prediction for proposed DSIs

Below are some examples of relations found with our method that were deemed to be of potential clinical interest by our experts. The first four examples in table 4 are discovered through DsGD pathway, while the last example is through DsGFGD pathway.

| Identified Interaction | Pathway | Clinical Context |
|---|---|---|
| glucosamine and prostaglandin | Glucosamine **INHIBITS** COX-2 **PRODUCES** prostaglandin | Glucosamine is commonly used as a supplement for osteoarthritis. Glucosamine is a natural compound which is found in cartilage. COX-2 or cyclooxygenase-2 is an inducible enzyme which has increased activity in inflamed joint tissues and increases prostaglandin E2 (PGE2) which is associated with inflammation and structural changes associated with arthritic disease [38]. Since the pathway shows glucosamine inhibiting COX-2 and thus reducing prostaglandin, this provides a mechanistic support for the use of glucosamine in treating arthritis. |
| grain alcohol and dinoprostone | grain alcohol **STIMULATES** Interleukin 1, Beta **STIMULATES** dinoprostone | Grain alcohol is noted to stimulate Interleukin 1, Beta which stimulates dinoprostone which has important effects in potentially inducing labor in pregnant women. Prior older studies have looked at the use of ethanol in stopping preterm labor [39] but other summaries have shown limited effectiveness vs bed rest alone and potential concerns regarding the side effects of alcohol on the fetus [40]. Given the association of alcohol with stimulation of IL-1Beta and dinoprostone, this would provide a mechanistic reason to avoid alcohol for treatment of preterm labor in addition to the adverse fetal effects. |
| Curcumin and Oxytocin | Curcumin **INHIBITS** Interleukin 1-beta **PRODUCES** Oxytocin | Curcumin is a yellow chemical produced by turmeric (Curcuma longa). It is sold as a herbal supplement and a culinary spice. Oxytocin is a medication used to contract the uterus to increase the speed of labor and to stop bleeding following delivery. We found that Interleukin-1-beta influences oxytocin signaling and production. Therefore, co-administration of both may reduce the effect of oxytocin on labor induction. |

| Micrococcus and vaccine | Micrococcus aureus<br><br>**STIMULATES**<br><br>Edodekin Alfa<br><br>**PRODUCES**<br><br>vaccine | Micrococcus aureus is also known as Staphylococcus Aureus and is associated with stimulating Edodekin Alfa which is also known as Interleukin 12. Edodekin Alfa is associated with the production of vaccine. The Edodekin Alfa has effects which can be used as an adjuvant to increase vaccine response and help produce an immunologic response to vaccine. Given that micrococcus aureus may be colonized in some individuals [41], it may be of interest to see if those who are colonized have greater vaccine responses. |
|---|---|---|
| dandelion and autophagies | Blowball (dandelion)<br>↓<br>**STIMULATES**<br>↓<br>AMPK<br>↓<br>**AUGMENTS**<br>↓<br>autophagies<br>↑<br>**DISRUPTS**<br>↑<br>Heme Oxygenase-1<br>↑<br>**STIMULATES**<br>↑<br>Australian tea tree oil | Dandelion (blowball) is associated with the stimulation of AMPK which is also associated with augmentation of autophagy. Autophagy has a role in metabolic adaptation, intracellular quality control, and renovation [42]. Australian tea tree oil stimulates Heme Oxygenase-1 which is then associated with disrupting autophagies. In looking at the two sets of mechanisms, there could be potential interactions between the use of dandelion supplements and Australian tea tree oil at the same time which could create potential issues if active in the same cellular components. |

Table 4: Proposed DSIs and expert evaluations

**5.4 Error Analysis**

We also conducted error analysis on false positives during evaluation. Errors stem from how terms are mapped to concepts when MetaMap tries to map text to a concept in our extended UMLS. An example of a mapping error due to MetaMap we observed is 'diet' mapping to 'Bill Henderson Protocol', a proposed diet to help fight cancer. Another involved contextual differences in the meaning of a word. For example, "contracted" identified in one context based on relationships with contraction of gel media in a laboratory experiment versus contraction of blood vessels.

Another source of error is with SemRepDS. The extraction of semantic relationships by SemRepDS has a precision of only 0.67. While filtering improved this to 0.91, there is still a non-trivial proportion of incorrect relations contained in SemMedDB-DS. As such, any of the proposed pathways that contained an incorrect relation would be rendered invalid based on our evaluation criteria.

The most common types of errors found in the sample of filtered relations were either SemRepDS missing the negation of a predicate or the improper attribution of an entity to a relation in a different clause. We include some illustrative examples of pathways that were incorrect due to a semantic error or extraction error. For example, the semantic relation "red wine prevents atherosclerosis" was extracted from "Red wine does not reduce mature atherosclerosis in apolipoprotein E-deficient mice." In this case, SemRepDS missed the negation of the predicate to produce an incorrect relation. In another example, "manganese prevents pre-eclampsia" extracted from "While calcium has been shown to reduce the risk of pre-eclampsia and maternal mortality, calcium, phosphorus, potassium, magnesium and manganese can have negative impacts on organoleptic properties, so many products tested have not included these nutrients or have done so in a limited way." Here, SemRepDS incorrectly associated manganese with "reduce the risk of" from the first clause containing 'pre-eclampsia'.

**5.5 Limitations and future work**
Available DS concepts and terms are inherently limited by those included in sources for iDISK, which imposes a limitation on the terms SemRepDS can identify. Thus, there might be abstracts that contain DS mentions and relations but cannot be identified by SemRepDS.

Another limitation is that the DSI discovery was performed on a graph generated by a subset of PubMed abstracts rather than all abstracts. This may have resulted in the unintentional exclusion of some relations and potential DSIs from our final list. This was due to computational limitations as processing the entirety of PubMed with SemRep takes around one month. There is additional work that can be done to improve the ability of SemRepDS to properly extract relations when negation is involved. We also plan to explore a larger knowledge graph that uses the entirety of SemMedBD and SuppKG.

We see that the BERT filtering significantly improved the precision of the relations, but there is still room for incorrect relations to have been included in SuppKG. This means that there are still potentially incorrect relations that were used in some of the pathways found with our patterns. The BERT filtering model substantially improved the precision of the semantic relations but there is room (e.g., increase the annotated data set, etc.) to further improve the quality of SuppKG with additional or refined filtering methods.

While the pathways we used produced novel interactions not found in the literature, there are other methods we would like to use. There are limitations to the use of rules based DSI discovery, namely that the pathways need to be decided on by experts and that the pathways are not precise since they don't use information contained in the knowledge graph aside from semantic types, resulting in a large volume of meaningless interactions. We will explore machine learning based methods such as standard embedding models (e.g. TransE), graph neural networks and transformer-based models which are not as interpretable as discovery patterns since they only give the initial and terminal nodes rather than a full pathway.

**6 CONCLUSION**
The UMLS Metathesaurus contains limited data specific to the growing DS domain. In this study, we demonstrate successful augmentation of SemRep with iDISK by using the Data File Builder to expand DS representation in the UMLS Metathesaurus. The resulting knowledge graph, SuppKG, was improved by training models to remove incorrect relations, thus reducing downstream error propagation. We also identified several novel DSIs not found in the literature and some that have potential clinical interest. These results all highlight the advantages and utility of augmenting the UMLS Metathesaurus with DS data.

**ACKNOWLEDGEMENTS**
This work was supported by the National Institutes of Health's National Center for Complementary and Integrative Health and the Office of Dietary Supplements grant number R01AT009457 (to RZ). The content is solely the responsibility of the authors and does not represent the official views of the National Institutes of Health.

**COMPETING INTERESTS**
None to declare.

**AUTHOR CONTRIBUTIONS**
DS, JV and RZ conceived the study design and wrote the initial draft of the manuscript. DS and JV carried out the experiments and produced the original draft of the manuscript. AB and YZ annotated the PubMed abstracts. TA and JB conducted the qualitative evaluations of literature-based discovery. HK, MF and HX were consulted on specifics of the experimental design. All authors contributed to the production of the final manuscript.